\documentclass[iop]{emulateapj}

\usepackage{graphicx}
\usepackage{amsmath}
\usepackage{appendix}
\usepackage{amssymb}
\usepackage{aas_macros}
\usepackage{natbib}
\usepackage{latexsym}
\usepackage{longtable}
\usepackage{threeparttable}
\usepackage{subfigure}
\usepackage[figoff]{figcaps}
\usepackage{ccaption}
\usepackage{hyperref}
\captiondelim{ }

\begin{document}

\pagestyle{plain}
\title{The Role of Kozai Cycles in Near-Earth Binary Asteroids}
\author{Julia Fang\altaffilmark{1} and Jean-Luc Margot\altaffilmark{1,2}}

\altaffiltext{1}{Department of Physics and Astronomy, University of California, Los Angeles, CA 90095, USA}
\altaffiltext{2}{Department of Earth and Space Sciences, University of California, Los Angeles, CA 90095, USA}

\begin{abstract}

We investigate the Kozai mechanism in the context of near-Earth
binaries and the Sun. The Kozai effect can lead to changes in
eccentricity and inclination of the binary orbit, but it can be
weakened or completely suppressed by other sources of pericenter
precession, such as the oblateness of the primary body. Through
numerical integrations including primary oblateness and 3 bodies (the
two binary components and the Sun), we show that Kozai cycles cannot
occur for the closely-separated near-Earth binaries in our sample. 
We demonstrate that this is due to pericenter precession around the oblate
primary, even for very small oblateness values.  Since the majority of
observed near-Earth binaries are not well-separated, we predict that
Kozai cycles do not play an important role in the orbital evolution of
most near-Earth binaries. For a hypothetical wide binary modeled after
1998~ST27, the separation is large at 16 primary radii and so the
orbital effects of primary oblateness are lessened. For this wide
binary, we illustrate the possible excursions in eccentricity and
inclination due to Kozai cycles as well as depict stable orientations
for the binary's orbital plane. Unstable orientations lead to
collisions between binary components, and we suggest that the Kozai
effect acting in wide binaries may be a route to the formation of
near-Earth contact binaries.

\end{abstract}
\maketitle

\section{Introduction} \label{introduction}

About 15\% of near-Earth asteroids (NEAs) larger than approximately
200 meters in diameter are in a binary configuration \citep{marg02,
prav06}. These NEA binaries have dynamical lifetimes on the order of a
few million years \citep{bott02} and are subject to a variety of
perturbations, including close scattering encounters with terrestrial
planets, radiative effects (called BYORP) from the Sun, tidal torques,
and the Kozai mechanism. Such perturbative effects may change the
orbital energy and angular momentum of the system, which can influence
the binary's orbital elements, including semi-major axis, eccentricity, and
inclination. We briefly discuss each of these perturbations next; their relevance
in explaining observed spin and orbital data of NEA binaries is
discussed in \citet{fang11ii}.

The effect and frequency of close planetary encounters are studied
by \citet{fang11i}, who found that approaches ($<$10
Earth radii) with Earth can occur for most observed NEA binaries on
1$-$10 million-year timescales. The radiative BYORP effect, while not
observationally verified to date, is only relevant for 
satellites with spin-orbit synchronization. This effect has been
theoretically found to be capable of modifying an NEA binary's
semi-major axis and eccentricity on fast ($\sim$10$^5$ years)
timescales
\citep{cuk05,cuk07,gold09,cuk10,mcma10b,mcma10a,stei11}. Tidal
evolution of NEA binaries has been previously studied (see
\citet{tayl11} and references therein), and can cause spin-orbital
synchronization as well as modify eccentricity
\citep[i.e.][]{gold63,gold09}. The Kozai mechanism \citep{koza62}, in
the context of NEA binaries, has not been fully studied and is the
focus of this paper. 

The Kozai mechanism is very relevant for many astrophysical triple
systems; examples include its influence on the stability for irregular
Jovian satellites with high inclinations \citep[e.g.][]{nesv03}, main
belt and trans-Neptunian binaries \citep{pere09}, asteroids and comets
due to Jupiter \citep[e.g.][]{koza62,thom96}, binary stars with
distant companions \citep{harr68,maze79,kise98,eggl01,eggl06,fabr07},
extrasolar planets with outlying perturbers
\citep{maze97,inna97,holm97,trem04,take05,fabr07,katz11,lith11,naoz11},
and binary supermassive black holes \citep{blae02}.

\def\arraystretch{1.4}
\begin{deluxetable*}{lrrrrrrrrrrr}
\tablecolumns{12}
\tablecaption{Sample of Near-Earth Binaries \label{binaries}}
\startdata
\hline \hline
\multicolumn{1}{c}{System} &
\multicolumn{1}{c}{$R_p$} &
\multicolumn{1}{c}{$M_p$} &
\multicolumn{1}{c}{$R_s$} &
\multicolumn{1}{c}{$M_s$} &
\multicolumn{1}{c}{$n$} &
\multicolumn{1}{c}{$a$} &
\multicolumn{1}{c}{$e$} &
\multicolumn{1}{c}{$R_{\rm Hill}$} &
\multicolumn{1}{c}{$n_{\odot}$} &
\multicolumn{1}{c}{$a_{\odot}$} &
\multicolumn{1}{c}{$e_{\odot}$} \\
\multicolumn{1}{c}{} &
\multicolumn{1}{c}{(km)} &
\multicolumn{1}{c}{(kg)} &
\multicolumn{1}{c}{(km)} &
\multicolumn{1}{c}{(kg)} &
\multicolumn{1}{c}{(deg d$^{-1}$)} &
\multicolumn{1}{c}{(km)} &
\multicolumn{1}{c}{} &
\multicolumn{1}{c}{(km)} &
\multicolumn{1}{c}{(deg d$^{-1}$)} &
\multicolumn{1}{c}{(AU)} &
\multicolumn{1}{c}{} \\
\hline
(276049) 2002~CE26\footnotemark[1] & 1.75 & 2.17 $\times$ $10^{13}$ & 0.150 & 1.37 $\times$ $10^{10}$ &  554.34 & 4.87   & 0.025  &	514	&  0.30  &	2.23	& 0.56 \\
2004~DC\footnotemark[2] 		   & 0.17 & 3.57 $\times$ $10^{10}$ & 0.030 & 1.96 $\times$ $10^{8\ }$&  372.93 & 0.75   & 0.30  &  	44	&  0.47  &	1.63	& 0.40 \\
(164121) 2003~YT1\footnotemark[3]  & 0.55 & 1.89 $\times$ $10^{12}$ & 0.105 & 1.32 $\times$ $10^{10}$ &  226.73 & 3.93   & 0.18  &	113	&  0.84  &	1.11	& 0.29 \\
(65803) Didymos\footnotemark[4]    & 0.40 & 5.24 $\times$ $10^{11}$ & 0.075 & 3.45 $\times$ $10^{9\ }$ &  724.38 & 1.18   & 0.04   &	109	&  0.47  &	1.64	& 0.38 \\
(35107) 1991~VH\footnotemark[5]    & 0.60 & 1.40 $\times$ $10^{12}$ & 0.240 & 8.93 $\times$ $10^{10}$ &  265.07 & 3.26   & 0.06   &	105	&  0.81  &	1.14	& 0.14 \\
Hypothetical Wide Binary	 	   & 0.42 & 7.73 $\times$ $10^{11}$ & 0.050 & 1.51 $\times$ $10^{9\ }$&  65.46  & 6.66   & 0.3    &	62	&  1.33  &	0.82	& 0.53
\enddata
\tablenotetext{}{This table consists of a subset of well-characterized NEA binaries whose 
uncertainty region for the orbital plane orientation includes the Kozai-acting regime 
(39.2$^{\circ}$ $<$ $i$ $<$ 140.8$^{\circ}$) for initially circular binaries.  In addition,
we list a hypothetical wide binary modeled after 1998~ST27, whose actual physical and orbital properties 
are not well known. For all entries, we list adopted values for primary radius $R_p$, primary mass $M_p$,
secondary radius $R_s$, 
secondary mass $M_s$, binary mean motion $n$, binary semi-major axis $a$, binary eccentricity $e$, 
the Hill radius $R_{\rm Hill}$ (beyond which the binary's components would be primarily 
orbiting the Sun instead of each other), heliocentric mean motion $n_{\odot}$, heliocentric 
semi-major axis $a_{\odot}$, and heliocentric eccentricity $e_{\odot}$. Binary parameters and 
uncertainties can be found in \citet{fang11ii}, and heliocentric parameters and uncertainties
can be found in the JPL Small Body Database.}
\footnotetext[1]{\citet{shep06}}
\footnotetext[2]{\citet{tayl08b}}
\footnotetext[3]{\citet{nola04}}
\footnotetext[4]{\citet{benn10}}
\footnotetext[5]{\citet{marg08,prav06}}
\end{deluxetable*}

In this paper, we consider the Kozai mechanism in terms of the
following triple system: an NEA binary, consisting of a massive
primary and less massive secondary, and the Sun as an outer perturber
(as in \citet{pere09}).  The Kozai mechanism is a {\em secular}
effect, i.e.\ the effect occurs on timescales that exceed the orbital
periods.  Analytical computation of secular effects traditionally
involve averaging quantities over a complete orbital cycle.  Under the
Kozai effect, the secondary's orbit with respect to the primary will
undergo coupled changes in eccentricity and inclination. We define the
inclination as the relative inclination between the binary's mutual
orbit and the binary's heliocentric orbit. For a circular heliocentric
orbit, the coupled oscillations in eccentricity $e$ and inclination
$i$ will conserve the quantity $\sqrt{1 - e^2}\cos i$ such that for
satellites in prograde orbits ($0^{\circ} \leq i < 90^{\circ}$), peaks
in eccentricity correspond to minima in inclination, and vice versa.
Satellites in retrograde orbits ($90^{\circ} \leq i \leq 180^{\circ}$)
will have eccentricity and inclination oscillate in the same
direction. In the absence of other perturbations, an initially
circular binary will undergo large Kozai cycles if a critical
inclination $i$ is met: 39.2$^{\circ}$ $<$ $i$ $<$
140.8$^{\circ}$. 
Initially eccentric binaries will undergo Kozai cycles over a wider
inclination range.  For an initially circular binary, the maximum
eccentricity $e_{\rm max}$ that can be induced by the Kozai mechanism
is
\begin{equation} \label{emax}
	e_{\rm max} = \sqrt{1 - \dfrac{5}{3}\cos^2 i_{\rm init}}
\end{equation}
where $i_{\rm init}$ is the initial inclination \citep[i.e.][]{inna97}. 
If $i_{\rm init} \sim$ 90$^{\circ}$, then the binary's eccentricity
will grow to $\sim$1 during a single Kozai oscillation. The
approximate oscillation timescale or Kozai period $P_K$ is on the
order of \citep{kise98}
\begin{equation} \label{kozaip}
	P_K = \dfrac{2 P_{\odot}^2}{3 \pi P} (1 - e_{\odot}^2)^{3/2} \dfrac{M_p + M_s + M_{\odot}}{M_{\odot}}
\end{equation}
where the binary's mutual orbit has period $P$, the primary's mass is
$M_p$, the secondary's mass is $M_s$, the Sun's mass is $M_{\odot}$,
the heliocentric orbital period is $P_{\odot}$, and the heliocentric
eccentricity is $e_{\odot}$. The Kozai effect due to the Sun for
binaries in the Solar System is much more pronounced for NEAs than for
main belt and trans-Neptunian objects because $P_K$ varies as the
square of the heliocentric orbital period $P_{\odot}$.

Among the population of well-characterized NEA systems compiled by
\citet{fang11ii}, we list in Table \ref{binaries} only binary systems
for which the uncertainty region in orbital plane orientation includes
the Kozai regime (39.2$^{\circ}$ $<$ $i$ $<$ 140.8$^{\circ}$) for 
initially circular binaries. This criterion rules out 2 triple systems,
2001~SN263 and 1994~CC, and 2 binary systems, 2000~DP107 and 1999~KW4,
for which the Kozai effect will not operate \citep{fang11ii}. We also
include a hypothetical wide binary modeled after 1998~ST27, which is
the widest NEA binary known so far but unfortunately has binary
orbital and physical parameters that are not well-determined. For the
hypothetical binary's heliocentric orbital elements listed in Table
\ref{binaries}, we use 1998~ST27's actual heliocentric orbital
elements, which are well-known.

The goal of this study is to investigate the relevance and effect of
Kozai cycles in the presence of other modulating perturbations. The
outline of this paper is as follows. In Section \ref{others}, we
describe other perturbations--the primary's oblateness, additional
satellites, and tidal effects--that may damp Kozai cycles and we find
that primary oblateness is dominant. In Section \ref{modulated}, we
perform numerical investigations of the Kozai effect modulated by
primary oblateness for all binaries in our sample, and determine their
excursions in eccentricity and inclination. In Section
\ref{discussion}, we discuss the implications of this work, including
the role of Kozai cycles in the evolution of most NEA binaries, constraints
on orbital plane orientations, how Kozai-induced instabilities end in
collisions, and the formation of contact binaries. We briefly summarize 
this study in Section \ref{conclusion}.

\section{Perturbations that affect Kozai cycles} \label{others}

Here, we consider effects that can weaken or completely suppress Kozai
cycles. The Kozai effect causes oscillations in eccentricity and
inclination, and systems in Kozai resonance can exhibit libration of
the argument of pericenter.  

Given that this effect is caused by the interaction
between the shape of the binary's mutual orbit and weak solar tides,
Kozai cycles can be easily suppressed by other weak perturbations
that contribute to apsidal (pericenter) precession in the binary.
If the argument of pericenter precesses too fast, libration of the
argument of pericenter is no longer possible, which inhibits the
Kozai process.

In the following subsections, we consider contributions to pericenter
precession from 3 effects: the primary's non-spherical shape leading
to a non-uniform gravitational field with a quadrupole moment ($J_2$)
representing the degree of oblateness, the presence of additional,
undetected satellites, and tidal bulges due to both the primary and
the secondary. These effects can potentially weaken or suppress Kozai
cycles, and we seek to determine their relative strengths.

\subsection{Primary Oblateness ($J_2$)}

We consider the contribution to pericenter precession caused by
primary oblateness. Many NEAs have non-spherical shapes, and the level
of oblateness can be described by a coefficient, $J_2$, as \citep{murr99}
\begin{align}
	J_2 = \dfrac{C - (A + B)/2}{M_pR_p^2} \approx \dfrac{C-A}{M_pR_p^2}
\end{align}
where $A$, $B$, and $C$ are the primary's moments of inertia. $M_p$
is the primary's mass and $R_p$ is the primary's equatorial radius.
The approximation is valid when $A \approx B$.

The $J_2$ coefficient is an indirectly observable quantity that can be 
detected by its non-Keplerian effects induced on orbiting satellites. 
Such oblateness-induced precession on the orbits of satellites can be
examined by the rate of change in the argument of pericenter $\omega$
\citep{vall01}:
\begin{equation} \label{j2}
	\dfrac{d\omega}{dt}_{\rm J_2} =  \dfrac{3}{2}\dfrac{nJ_2}{(1-e^2)^2}\left(\dfrac{R_p}{a}\right)^2\left(\dfrac{5}{2}\cos^2I-\dfrac{1}{2}\right) 
\end{equation}
where $n$ is the binary's mean motion, $e$ is the eccentricity, $R_p$
is the primary's radius, $a$ is the semi-major axis, and $I$ is the
binary's orbital inclination relative to the primary's equator.
The $J_2$ effect
is more relevant for asteroid binaries than their trans-Neptunian
counterparts because asteroid binaries tend to be separated by several
primary radii and trans-Neptunian binaries are typically much
wider. The orbits of close-in satellites will be more perturbed by an
oblate primary than the orbits of distant satellites, since Equation
\ref{j2} shows that precession
due to $J_2$ varies inversely as distance squared.

The primary's $J_2$ value for most NEA binaries is unknown; we
calculate a range of values for apsidal precession due to $J_2$
coefficients ranging from 0.001 to 0.1. A few well-characterized
NEA systems have known shapes and $J_2$ values, including
1999~KW4 \citep{ostr06} with a primary $J_2$ of $\sim$0.06 and 
1994~CC \citep{broz11} with a primary $J_2$ of $\sim$0.01.
In our calculations, 
we assume that the satellite is orbiting in the equatorial plane of the primary
\citep[support for this assumption comes from the generally accepted
  rotational fission formation model;][]{marg02,prav06,wals08}, and we
use nominal values of their current separations and
eccentricities. These numbers are presented in Table \ref{rates}.

\subsection{Additional Satellite}

We examine if the presence of an additional satellite 
(presumably undetected) in the asteroid system can suppress Kozai
oscillations by causing the known satellite's orbit to precess. It is
known that the presence of larger, detectable satellites can easily
suppress Kozai cycles, as shown through numerical simulations by
\citet{fang11} for NEA triple systems 2001~SN263 and 1994~CC and as
discussed by \citet{rago09} for trans-Neptunian triple Haumea. Here,
we investigate if a small, unobserved satellite can also damp Kozai
cycles. 

Since most NEA binaries are discovered by planetary radar, we adopt a
fairly common value of the spatial resolution in radar images (15 m)
as the typical radius of the additional satellite in our study.  This
is adopted for convenience and does not represent the finest spatial
resolution available nor the radar detectability threshold.  If a
small, undetectable satellite can damp Kozai cycles, then it may be
responsible for the survival of NEA systems that would otherwise
undergo high oscillations in eccentricity and inclination.

In the presence of an additional satellite, both satellites will
undergo time-averaged, secular changes in their orbital elements. For
a coplanar system, their argument of pericenter rates are
\citep{mard07}
\begin{align} \label{body1}
	\dfrac{d\omega_{s1}}{dt}_{\rm sat} &= \dfrac{3}{4} n_{s1} \left(\dfrac{M_{s2}}{M_p} \right) \left(\dfrac{a_{s1}}{a_{s2}}\right)^3 (1-e_{s2}^2)^{3/2} \\
	 \nonumber & \times \left[ 1 - \dfrac{5}{4} \left(\dfrac{a_{s1}}{a_{s2}}\right) \left(\dfrac{e_{s2}}{e_{s1}}\right) \dfrac{\cos(\omega_{s1} - \omega_{s2})}{1 - e_{s2}^2} \right]
\end{align}

\begin{align} \label{body2}
	\dfrac{d\omega_{s2}}{dt}_{\rm sat} &= \dfrac{3}{4} n_{s2} \left(\dfrac{M_{s1}}{M_p}\right) \left(\dfrac{a_{s1}}{a_{s2}}\right)^2 (1-e_{s2}^2)^{-2} \\
	 \nonumber & \times \left[ 1 - \dfrac{5}{4} \left(\dfrac{a_{s1}}{a_{s2}}\right) \left(\dfrac{e_{s1}}{e_{s2}}\right) \dfrac{(1+4 e_{s2}^2)}{(1-e_{s2}^2)} \cos(\omega_{s1} - \omega_{s2}) \right]
\end{align}
and these equations are given to fourth power in $a_{s1}/a_{s2}$ and
first order in $e_{s1}$.
The subscripts $s1$ and $s2$ represent the inner and outer
satellites, respectively, and the subscript $p$ is for the
primary. The equations also include the mean motion $n$, mass $M$,
semi-major axis $a$, and eccentricity $e$. These equations are valid
for $M_{s1} << M_p$, but there are no restrictions on $M_{s2}$.

We calculate the apsidal rates for each binary in Table
\ref{binaries}, assuming a coplanar system. For each binary, the
additional satellite is given a radius of 15 meters 
and a typical rubble pile density of 2 g cm$^{-3}$ with a circular
orbit. For all binaries except the hypothetical wide binary modeled
after 1998~ST27, we treat the observed satellite as the inner
satellite and the additional, undetected satellite as the outer
satellite with a semi-major axis of 15 primary radii (which is a
typical separation for the outer satellite in an NEA triple, based
on a sample of 2 known NEA triples). 
For the hypothetical wide binary, whose satellite is located at 6.66
km or 16 primary radii, we treat this satellite as the outer satellite
and the undetected satellite as the inner satellite with a semi-major
axis of 4 primary radii. In all cases, we use the observed satellite's
current separation and eccentricity. Our calculated apsidal precession
rates are shown in Table \ref{rates}, 
which represent the pericenter precession due to a 30~m diameter
satellite.

\def\arraystretch{1.4}
\begin{deluxetable}{lrrr}
\tablecolumns{4}
\tablecaption{Pericenter Precession Rates \label{rates}}
\startdata
\hline \hline
\multicolumn{1}{c}{System} &
\multicolumn{3}{c}{Pericenter Precession (deg d$^{-1}$):} \\
\multicolumn{1}{c}{} &
\multicolumn{1}{c}{$\dot{\omega}_{\rm J_2}$} &
\multicolumn{1}{c}{$\dot{\omega}_{\rm sat}$} &
\multicolumn{1}{c}{$\dot{\omega}_{\rm tides}$} \\
\hline
2002~CE26 & 0.22$-$22  	    &  $<$3.5e-6 & 5.0e-7 $-$ 3.1e-5 \\
2004~DC   & 0.069$-$6.9     &  $<$5.6e-3 & 5.7e-8 $-$ 8.2e-5 \\
2003~YT1  & 0.014$-$1.4     &  $<$2.8e-4 & 8.6e-9 $-$ 1.0e-6 \\
Didymos   & 0.25$-$25	    &  $<$2.2e-4 & 1.3e-6 $-$ 3.0e-4 \\
1991~VH   & 0.027$-$2.7     &  $<$1.9e-4 & 3.3e-7 $-$ 8.3e-6 \\
HWB*      & 0.00094$-$0.094 &  $<$1.4e-4 & 1.5e-11$-$ 6.7e-9
\enddata
\tablenotetext{}{Rates for the argument of pericenter $\omega$ are
  calculated for each binary and for three main sources of pericenter
  precession: primary's $J_2$, an additional satellite, and tidal
  bulges raised on the primary and secondary. 
  A range of rates is given for the effects of oblateness, 
  corresponding to $J_2$ values from 0.001 to 0.1.
  A range of rates is given for the effects of tides, 
  corresponding to two different tidal Love
  number models \citep{gold09,jaco11}.
  Precession due to an additional satellite assumes a size of 30~m. \\
*HWB = hypothetical wide binary modeled after 1998~ST27}
\end{deluxetable}

\subsection{Tidal Bulges}

For completeness we investigate tidal bulges raised on both the
primary and secondary, although we anticipate their contribution to
pericenter precession to be small.  The argument of pericenter rate is
\citep[e.g.][]{ster39, fabr07, baty11}
\begin{align} \label{tides}
	\dfrac{d\omega}{dt}_{\rm tides} &= \dfrac{15}{16} n \left[\dfrac{8+12e^2+e^4}{(1-e^2)^5}\right] \\
	\nonumber & \times \left[ k_p \left(\dfrac{R_p}{a}\right)^5 \left(\dfrac{M_s}{M_p}\right) + k_s \left(\dfrac{R_s}{a}\right)^5 \left(\dfrac{M_p}{M_s}\right) \right]
\end{align}
where the first term corresponds to the primary and the second term
corresponds to the secondary. The equation also includes the mean
motion $n$, eccentricity $e$, radius $R$, semi-major axis $a$, mass
$M$, and tidal Love number $k$. Subscripts $p$ and $s$ stand for the
primary and secondary, respectively.

The tidal Love number is a poorly known quantity for small rubble pile
asteroids, and there are currently two different rubble pile models
that describe the Love number's dependence on size (or radius
$R$). \citet{gold09} give the relation $k_{\rm rubble} \sim 10^{-5}$
($R$/1~km), and \citet{jaco11} find that $k_{\rm rubble} \sim 2.5
\times 10^{-5}$ (1~km/$R$). Using both tidal Love number models and
the binaries' current separations and eccentricities, we calculate a
range of possible apsidal precession rates due to tidal bulges for each NEA
binary in Table \ref{rates}.

Examination of the values in Table \ref{rates} indicates that
even the smallest amount of primary oblateness ($J_2$ of 0.001)
will dominate over all other perturbations; the contribution
from each perturbation is quantified in Table \ref{rates} 
for all NEA binaries in our sample. Therefore, the numerical
integrations described in the next section only include the effect of
primary oblateness out of the 3 sources of pericenter precession
considered here.

\section{Numerical Investigation} \label{modulated}

Through numerical simulations, we explore the excursion in a binary's
eccentricity and inclination due to 3-body effects, such as Kozai
cycles induced by the Sun, and we include perturbations due to primary
oblateness.  We use a Bulirsch-Stoer algorithm from {\em Mercury}
\citep{cham99} and our system is composed of 3 bodies: the Sun and
primary and secondary components of the binary. For each binary in our
investigation (Table \ref{binaries}), our integration time covers at
least 10 Kozai oscillation periods with a timestep that is 1/50$^{\rm
  th}$ of the binary's mutual orbital period. Initial conditions for
each binary's heliocentric semi-major axis and eccentricity as well as
starting values for the masses, separations, and eccentricities of the
binary components are taken from their known, observed values (Table
\ref{binaries}).

For each binary in our sample, we perform an ensemble of simulations.
In all simulations we assume that the binary's mutual orbit is in the
primary's equatorial plane.  We sample a range of $J_2$ values to
approximate the primary's non-spherical shape, a range of inclinations
between the binary's mutual and heliocentric orbits, and a range of
values for the binary's argument of pericenter; none of these
parameters are known for the binaries in our sample.  Our choice of
$J_2$ values includes 0.001, 0.005, 0.01, 0.05, and 0.1.  Currently,
the best-characterized NEA binary is 1999 KW4 \citep{ostr06} with
detailed shape and orbital information, and its primary has a $J_2$
value of $\sim$0.06. Another NEA system with known primary shape is
1994~CC \citep{broz11}, with a primary $J_2$ of $\sim$0.01. Our range
of inclinations spans angles in the strongest Kozai-operating regime
of 40$^{\circ}$$-$90$^{\circ}$ with 10$^{\circ}$ increments; these
inclinations represent satellites in prograde orbits.
Identical behavior, but mirrored across 90$^{\circ}$, would be seen
for inclinations greater than 90$^{\circ}$ representing retrograde
orbits.  We sample arguments of pericenter from 0$^{\circ}$ to
90$^{\circ}$ with 30$^{\circ}$ increments, as similar behavior is
repeated for each 90$^{\circ}$ quadrant.

\begin{figure*}[htb]
	\centering
	\mbox{\subfigure{\includegraphics[height=2.55in]{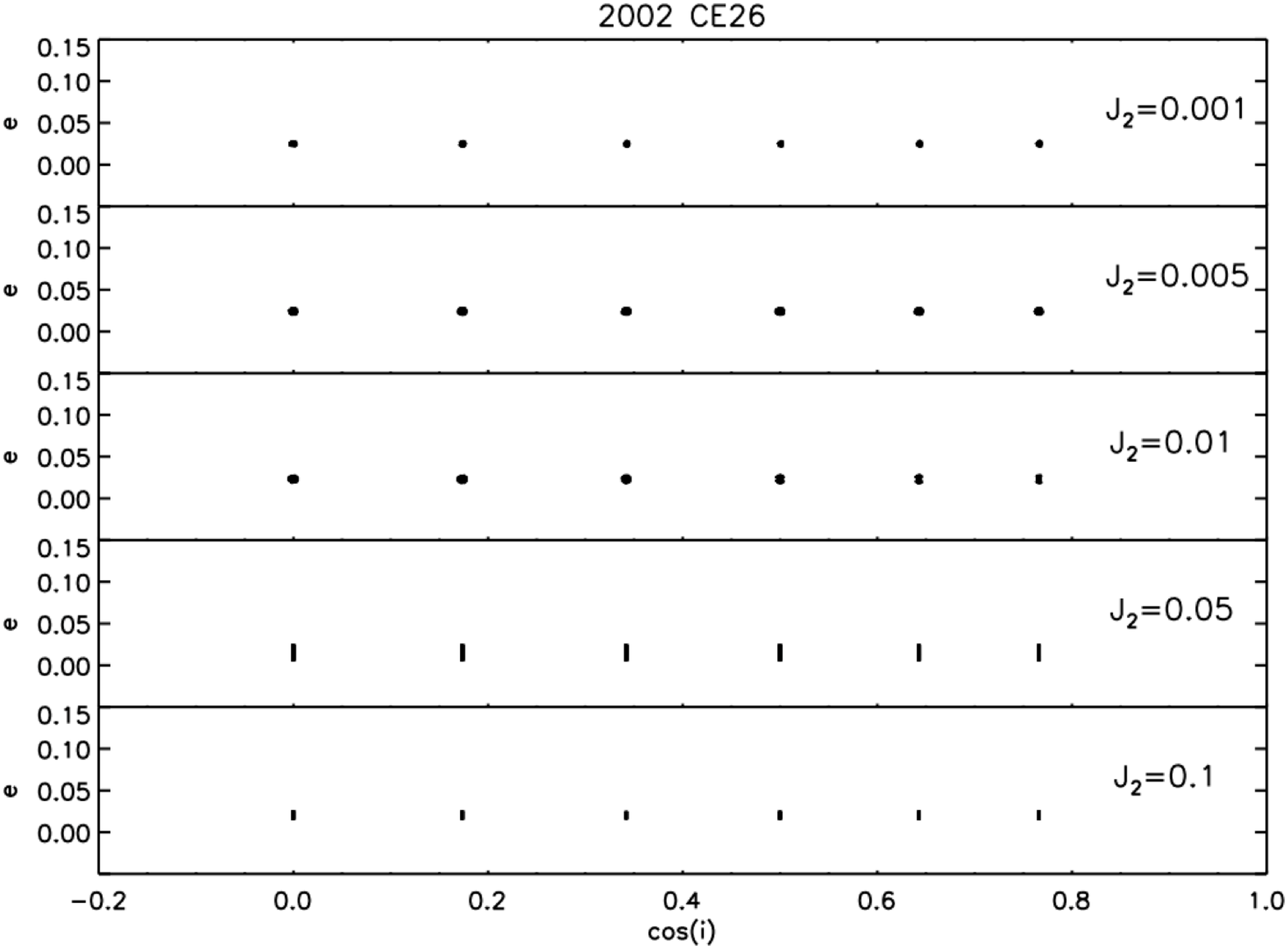}}\quad
	\subfigure{\includegraphics[height=2.58in]{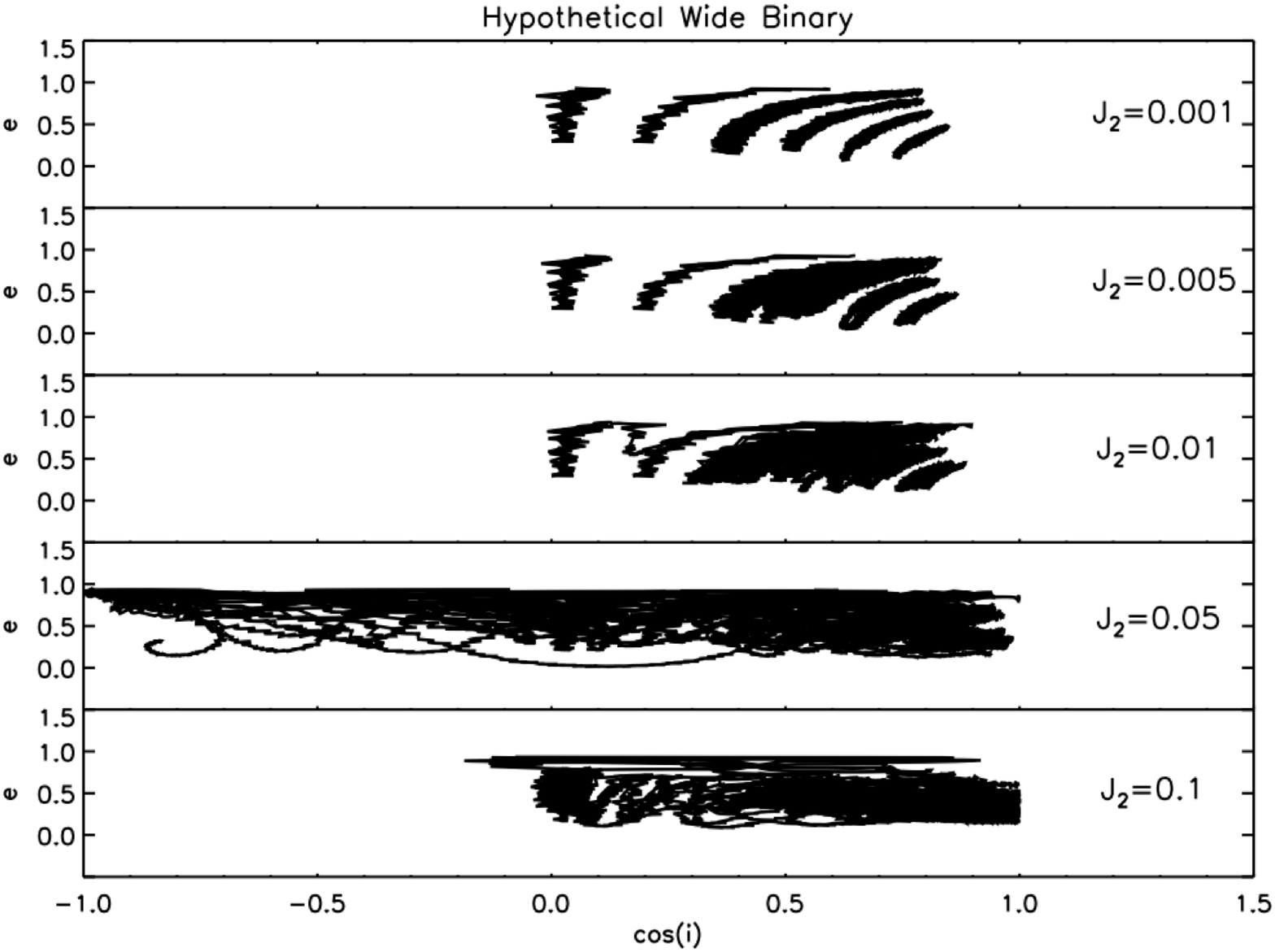}}}
	\caption{Results from all numerical simulations for 2002~CE26
          (left) and for a hypothetical wide binary modeled after
          1998~ST27 (right) are shown in these stacked plots with
          panels representing different $J_2$ values for the
          primary. The y-axis shows excursions in eccentricity $e$
          space and the x-axis illustrates the ranges of inclination
          $i$. In each panel, starting conditions include the observed
          separation and eccentricity as well as various values for
          the inclination and the argument of
          pericenter.  \label{stack}}
\end{figure*}

Our findings are as follows. Due to a non-zero $J_2$, binaries 2002
CE26, Didymos, 2004 DC, and 1991 VH do not exhibit any expected Kozai
behavior such as libration of the argument of pericenter or large
excursions in eccentricity and inclination. Instead, the argument of
pericenter circulates quickly and there are rapid, small-amplitude
oscillations in eccentricity and inclination. Typical behavior for
these binaries that do not exhibit expected Kozai cycles are
illustrated in Figure \ref{stack}'s left panel. This figure displays
the results from all numerical simulations for 2002~CE26 by showing
the eccentricity and inclination excursions. Moreover, if Kozai cycles
are present, then in the absence of other perturbers we expect an
initially circular binary to have an eccentricity increase (see
Equation \ref{emax}) of $\sim$0.15 for $i_{\rm init}$ of 40$^{\circ}$,
$\sim$0.56 for $i_{\rm init}$ of 50$^{\circ}$, $\sim$0.76 for $i_{\rm
  init}$ of 60$^{\circ}$, $\sim$0.90 for $i_{\rm init}$ of
70$^{\circ}$, $\sim$0.97 for $i_{\rm init}$ of 80$^{\circ}$, and
1 for $i_{\rm init}$ of 90$^{\circ}$. Comparison between these
values and Figure \ref{stack}'s left panel for 2002~CE26, an initially
near-circular binary, provides additional evidence that the expected
Kozai cycles are not present. For binary 2003~YT1, there are signs of
Kozai cycles only with a $J_2$ as low as 0.001, where there are
significant excursions in eccentricity and inclination; larger $J_2$
values suppress any Kozai oscillations.

Kozai cycles may induce collisional disruptions between the primary
and secondary if the eccentricity grows large enough that the
pericenter approaches one primary radius.  This is more likely to
occur at high inclinations (Equation \ref{emax}).  For binaries
2002~CE26, 2004~DC, Didymos, 1991~VH, and 2003~YT1, collisional
disruptions are expected at high inclinations yet not observed in any
of our simulations for the entire explored range of $J_2$ values.
None of the simulations for each of these binaries show any disruptions
(i.e.~collisions or ejections) during the nominal integration time,
contrary to expected Kozai behavior that leads to high eccentricities
within one cycle.  The lack of disruptions in non-zero $J_2$ cases
provides further confirmation that expected Kozai behavior is not
present when we include the effects of primary oblateness.

\begin{figure}[htb]
	\centering
	\includegraphics[scale=0.35]{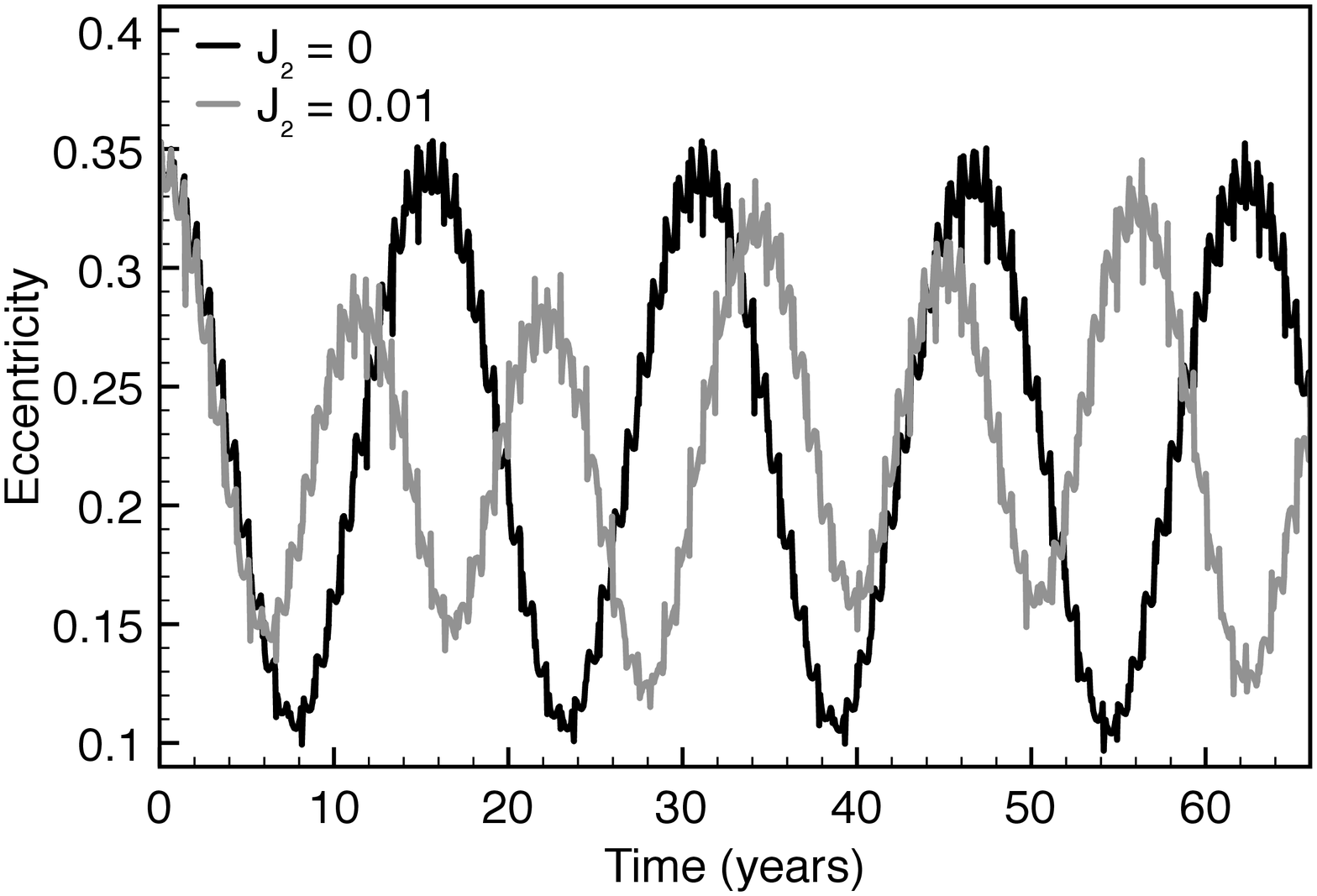}
	\includegraphics[scale=0.35]{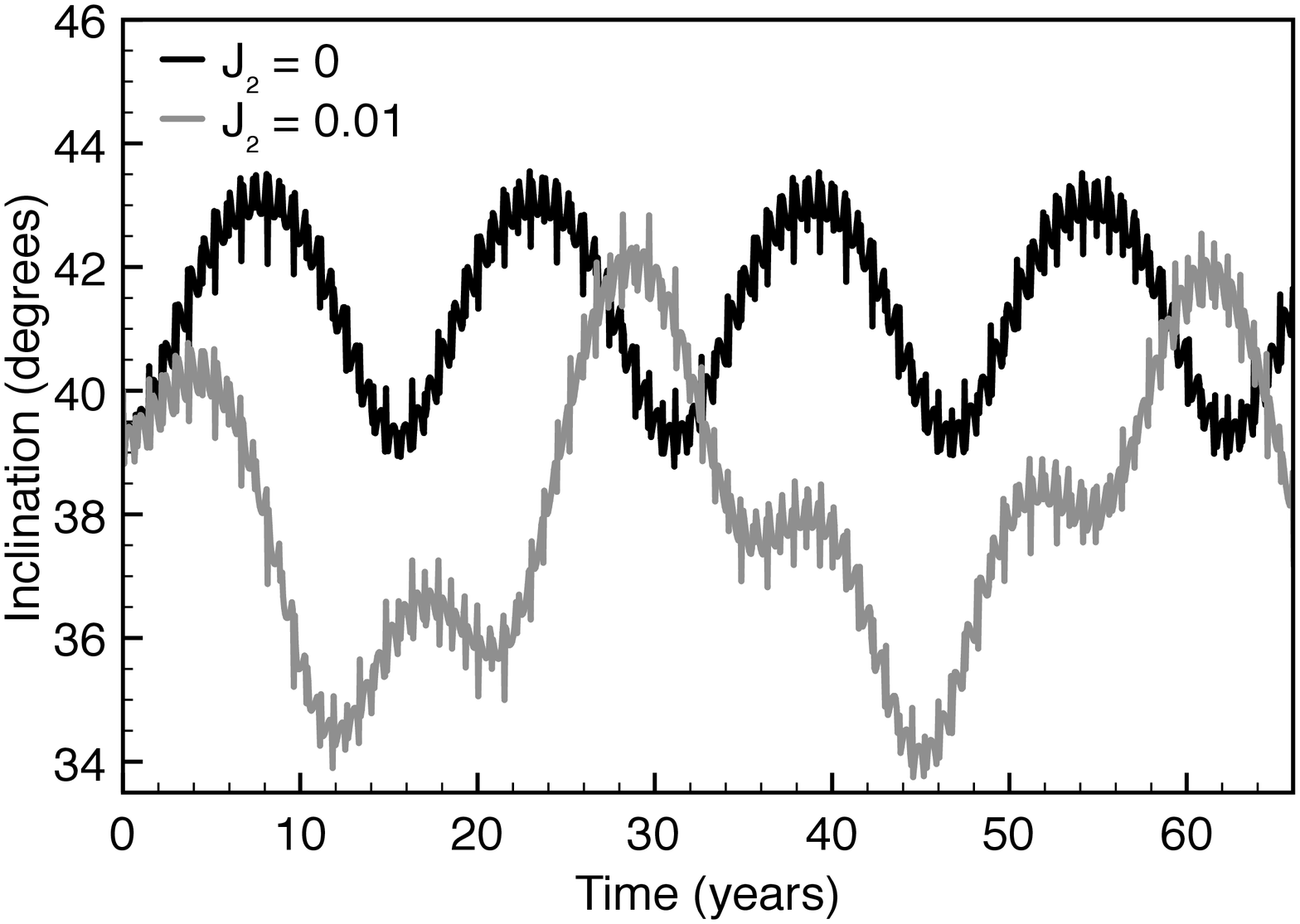}
	\caption{Here is a comparison between pure Kozai cycles ($J_2
          = 0$) and modulated Kozai cycles ($J_2 = 0.01$) for a
          hypothetical wide binary modeled after 1998~ST27. The
          starting conditions for this particular simulation are as
          follows: eccentricity is 0.3, inclination is 40$^{\circ}$,
          and the argument of pericenter is 90$^{\circ}$.
\label{st27_ei}} 
\end{figure}

For the hypothetical wide binary modeled after 1998~ST27, Kozai cycles
are present with the complete range of $J_2$ values sampled here.  In
the case of high $J_2$ values, the system exhibits wide, overlapping
excursions in eccentricity and inclination (Figure \ref{stack}'s
right panel). These Kozai cycles are modulated by a non-zero $J_2$; Figure
\ref{st27_ei} illustrates a comparison of pure Kozai cycles vs.~Kozai
cycles under the influence of $J_2$. This hypothetical binary exhibits
Kozai behavior as opposed to the other binaries in our sample because
it is the most widely-separated system with its satellite stationed at
a relatively far separation of 16 primary radii. Therefore, solar
perturbations are stronger than the effects induced by primary
oblateness. For this hypothetical wide binary only, we
continue integrating its ensemble of numerical simulations to 10$^5$
years to obtain more accurate disruption statistics.
We find that Kozai-induced disruptions occur in $\sim$44\% of all cases
examined here, with collisions and not ejections as the only observed
disruption outcomes.  See Section \ref{discussion} for discussion
regarding such instabilities.

These numerical integrations suggest that the evolution of most observed NEA
binaries are dominated by primary oblateness rather than the Kozai
effect. This conclusion is also supported by an analytical comparison
between the two effects. For all well-characterized binaries,
\citet{fang11ii} calculated the critical semi-major axis separating
the influence regions of primary oblateness and solar dynamics as
\citep{nich08}:
\begin{equation} \label{nichoeqn}
	a_{\rm crit} = \left( 2 J_2 \dfrac{M_p}{M_{\odot}} R_p^2 a_{\odot}^3 \right)^{1/5}
\end{equation}
where $J_2$ represents the extent of primary oblateness, $M_p$ is the
primary's mass, $M_{\odot}$ is the Sun's mass, $R_p$ is the primary's
radius, and $a_{\odot}$ is the heliocentric semi-major
axis. \citet{fang11ii} found that the range of allowable $a_{\rm
  crit}$ values corresponding to $J_2$ values ranging from 0.001 to
0.1 were greater than observed semi-major axes for all
well-characterized NEA binaries; they are oblateness-dominated. We
also calculate $a_{\rm crit}$ for the hypothetical wide binary modeled
after 1998~ST27. For this hypothetical binary, $a_{\rm crit}$ ranges
from 3.02 $-$ 7.60 km and so could be larger or smaller than the
adopted semi-major axis of 6.66 km.  Therefore, this hypothetical
binary could be in the oblateness-dominated or Kozai-dominated
regime. This analysis shows general consistency with the results from
numerical simulations; binaries predicted to be oblateness-dominated
by Equation \ref{nichoeqn} do not show expected Kozai oscillations in
numerical simulations, and the hypothetical wide binary predicted to
be either oblateness or solar dominated does show strong signs of
Kozai cycles modulated by $J_2$. Accordingly, tightly-bound binaries
(those that exhibit no signs of the Kozai effect) are strongly
affected by primary oblateness and loosely-bound binaries (showing
Kozai oscillations) are less perturbed by the far-away primary's
oblateness.

\section{Discussion} \label{discussion}

In this section, we discuss each of the following in turn: the
relevance of Kozai cycles for observed NEA binaries, orbit
orientation constraints for wide binaries that can undergo Kozai
oscillations, collisions as the only disruption outcome due to
the Kozai effect for typical NEA binaries, and Kozai
cycles as a possible route to the formation of contact binaries.

\subsection{General Population of Observed NEA Binaries}

We extrapolate from our simulation results to the observed NEA binary
population. Numerical integrations in Section \ref{modulated}
demonstrated that 4 out of 5 well-characterized NEA binaries in our
sample (Table \ref{binaries}) showed no signs of Kozai cycles and the
exception, 2003~YT1, only showed evidence of Kozai cycles when the
primary had a very minimal level of oblateness ($J_2 = $0.001). All 5
of these well-characterized binaries have semi-major axes less than 8
primary radii. Only the hypothetical wide binary (at 16 primary radii)
modeled after 1998~ST27 exhibited signs of Kozai cycles at a range of
$J_2$ values from 0.001$-$0.1. The effects of primary oblateness are
strongest for close-in binaries (pericenter precession due to $J_2$
increases as the semi-major axis squared; Equation \ref{j2}), where an
oblate primary can cause the satellite's orbital plane to precess fast
enough to thwart any Kozai oscillations.

If we consider all well-characterized NEA binaries as compiled by
\citet{fang11ii}, none of them have separations greater than 8 primary
radii. In fact, the semi-major axes of all of these well-characterized
NEA binaries are within the range of semi-major axes of the binaries
in our sample, which did not show Kozai cycles at a range of plausible
$J_2$ values. If we consider this list of well-characterized NEA
binaries to be representative of the observed population of NEA
binaries, then typical NEA binaries do not appear to be affected by
Kozai cycles in their orbital evolution.  We point out a mild
selection effect in that most well-characterized NEA binaries have
made close enough approaches to Earth to be detected by radar, and
this subset of the population may have fewer wide binaries than the
general population.  One could argue that we mostly observe binaries
immune from Kozai effects because the Kozai-acting binaries have
disrupted on short Kozai timescales (see following subsections on
disruptions and contact binary formation).  However, our simulations
show that the requirements on primary oblateness and component
separation are quite stringent, and we conclude that Kozai cycles are
unlikely to be an important effect in most NEA binaries.

\subsection{Constraints on Orbit Pole Orientations}

Orbit pole orientations can be constrained for binaries that undergo
Kozai cycles. In Section \ref{modulated}, we explored a hypothetical
wide binary modeled after 1998~ST27 that exhibited signs of Kozai
oscillations and disruptions, and in this subsection we illustrate
this binary's stable orbit pole orientations. Stable orbit poles are
defined as initial orientations that do not end in a collision after
10$^5$ years of numerical integrations; these results are based on the
ensemble of simulations performed in Section \ref{modulated} as well
as additional simulations to sample the full range of
inclinations. The range of inclinations corresponding to stable
binaries allows us to calculate the corresponding J2000 ecliptic
coordinates (latitude $\beta$, longitude $\lambda$) of binary orbits
for stable binaries. Since the hypothetical wide binary under
consideration is modeled after 1998~ST27, we use 1998~ST27's
well-known heliocentric orbit poles (ecliptic longitude of the
ascending node$=$197.5842$^{\circ}$, ecliptic
inclination$=$21.05458$^{\circ}$) for our calculations.

\begin{figure}[htb]
	\centering
	\includegraphics[scale=0.5]{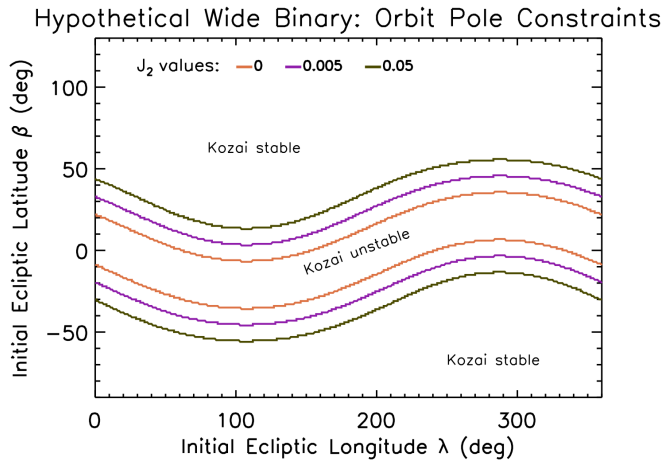}
	\caption{This stability map shows constraints on initial orbit
          pole orientations for a hypothetical wide binary modeled
          after 1998~ST27. There are 2 regions: ``Kozai stable''
          refers to regions where at least 50\% of simulations
          resulted in stable binaries, and ``Kozai unstable'' refers
          to regions where more than 50\% of simulations resulted in
          unstable binaries. Intermediate lines of varying colors
          (representing different $J_2$ values) separate ``Kozai
          stable'' and ``Kozai unstable'' regions. Stable orbit orientations 
	cover roughly $\sim$75\%, $\sim$50\%,
	and $\sim$45\% of the celestial sphere for $J_2$ values of 0, 0.005, and
	0.05, respectively. Stability is
          defined as binaries with no disruptions after 10$^5$ years
          in the ensemble of numerical integrations (Section
          \ref{modulated}).
\label{orbitpoles}} 
\end{figure}

In Figure \ref{orbitpoles}, we map out constraints on the orbit
orientations for this hypothetical wide binary, showing results for 3
different $J_2$ values for the primary: 0, 0.005, and
0.05. Color-coded lines (for different $J_2$ values) separate the
``Kozai stable'' and ``Kozai unstable'' regions. Regions where at
least half of simulations resulted in stable binaries are called
``Kozai stable,'' and regions where over half of simulations resulted
in unstable binaries are called ``Kozai unstable.'' This orbit
orientation map provides constraints on allowable orientations because
binaries that will disrupt under the effect of Kozai are unlikely to
be observed with orbit ecliptic coordinates in the unstable regions of
Figure \ref{orbitpoles}.  Most Kozai-induced disruptions occurred
within 100 years, shorter than evolutionary timescales due to tides,
BYORP, and close planetary encounters. Accordingly, for this
hypothetical wide binary we can predict that its orbit does not lie in
the region bounded by the color-coded $J_2$ lines.
This analysis is performed here for this hypothetical wide binary and
can similarly be applied to other binaries that can be affected by
Kozai perturbations.  We note that a different binary (presumably with
a different heliocentric orbital pole) would result in a different
layout of stable and unstable zones (i.e.~stability islands).

Figure \ref{orbitpoles} also shows how different $J_2$ values can
affect the stability map, discussed here for the hypothetical
wide binary. For discrete $J_2$ values of 0, 0.001,
0.005, 0.01, and 0.05, we find that an increasingly larger $J_2$ value
results in an increase in the number of unstable orbit orientations in
our simulations. These higher $J_2$ values cause collisions that
can occur at a larger range of initial inclinations by increasing
the maximum eccentricity acquired during a sequence of Kozai cycles.
A larger maximum eccentricity therefore decreases the pericenter
distance to allow for more frequent collisions \citep[e.g.][]{nesv03}.
We do not continue to observe this trend at high $J_2$ values greater
than $\sim$0.05, where substantial primary oblateness prevents
additional unstable orientations from forming.

\subsection{Kozai-Induced Disruptions End in Collisions}

Here we discuss how disruptions due to the Kozai effect will result in
collisions for typical NEA binaries. We only consider potential cases
where the Kozai effect operates and can produce instabilities, and we
assume these instabilities occur on much shorter timescales (on the
order of tens to hundreds of years) compared to other perturbations
such as planetary encounters, tidal evolution, and BYORP.
We define disruptions as
dynamical instabilities due to the Kozai effect, whose outcomes
include (1) collisions between binary components and (2) ejections
where the binary becomes unbound. The Kozai
effect is capable of increasing a binary's eccentricity, and very high
eccentricities can cause the binary to become unbound or cause a
collision between the binary components. An ejection occurs when an
orbit's apocenter $Q$, which is related to the semi-major axis $a$ and
the eccentricity $e$ as $Q = a(1+e)$, grows to a distance greater than
the binary's Hill radius $R_{\rm Hill}$. A collision occurs when an
orbit's pericenter $q$, where $q = a(1-e)$, is less than the primary
radius $R_p$. $R_{\rm Hill}$ and $R_p$ are known quantities given in
Table \ref{binaries}. Therefore, for given $R_{\rm Hill}$, $R_p$, and
$a$, whether a gradually increasing eccentricity (i.e.~due to the
Kozai effect) causes a binary to undergo a collision or ejection is
solely determined by the value of the eccentricity.

For all NEA binaries studied here (Table \ref{binaries}), values for
their maximum $Q$ (by assuming an $e$ of 1) are much smaller than
their Hill radii. In order for $Q$ to be as large as the Hill radius,
if we assume a maximum $e$ of 1 then any binary's semi-major axis
needs to be at least half as large as the Hill radius in order for an
ejection to occur. Thus, for any typical binary with a semi-major axis
smaller than 0.5$R_{\rm Hill}$, Kozai cycles will not cause
ejections. On the other hand, collisions will occur at an eccentricity
less than 1.  Collisions will occur (by setting $q = R_p$) at the
following $e$ values for the binaries in our sample: 0.64 for
2002~CE26, 0.77 for 2004~DC, 0.94 for the hypothetical wide binary,
0.86 for 2003~YT1, 0.66 for Didymos, and 0.82 for 1991~VH.  The
specific case of the hypothetical wide binary is most interesting
because we have already shown that the other binaries do not generally
undergo Kozai cycles, let alone Kozai-induced instabilities. But for
any wide NEA binary such as the hypothetical case examined here, if
Kozai cycles are acting with sufficiently high inclinations to induce
high eccentricities, collisions between binary components are the only
possible instability outcome. The binary's pericenter will shrink
below the primary radius before its apocenter can increase beyond the
Hill radius and become unbound.  Since collisions are the only
instability outcome, this means that the Kozai effect cannot directly
form asteroid pairs.  It is possible that the Kozai effect may
indirectly form asteroid pairs if a collision occurs but the
components separate again and become unbound, or if the binary
fortuitously makes a close planetary encounter during the short Kozai
timescale.

For Kozai cycles in the absence of other perturbations, these
collisions occur when the starting inclination results in a maximum
eccentricity (Equation \ref{emax}) that is greater than the limiting
eccentricities listed in the previous paragraph. Thus, for these
starting inclinations, collisions will occur over the course of one
Kozai cycle and repeated cycles will not occur. These disruption
timescales are fast; for the hypothetical binary modeled after
1998~ST27, our longer-term (10$^5$ years) simulations show that from a
starting eccentricity of 0.3, collisions occur ranging from $\sim$3
years to $>$10$^4$ years later with the majority of collisions
occurring within 100 years.

\subsection{Formation of Contact Binaries}

We discuss contact binary formation from large-amplitude Kozai
oscillations. For wide binaries such as 1998~ST27 that can potentially
undergo Kozai cycles even in the presence of primary oblateness, high
starting inclinations can lead to high eccentricities (Equation
\ref{emax}). During a single Kozai oscillation, these high
eccentricities can drive the orbital pericenter distance to very low
values and can cause the binary components to collide (see previous
section for discussion on collisions). If a collision does not occur,
tidal friction may play an important role and decrease the semi-major
axis as well as the eccentricity. When the pericenter distance
decreases due to high eccentricities, tides can more efficiently
circularize the orbit by dissipating orbital energy during each
pericenter passage \citep[e.g.][]{pere09,fabr07}. This leads to a
circular orbit and a smaller semi-major axis.

If Kozai cycles and possibly tidal effects can cause binary components
to come into contact with each other, this may be a mechanism to create 
near-Earth contact binaries, which constitute $\sim$10\% of NEAs
larger than 200~m in diameter \citep{benn06}.  If this process occurs,
we would expect the observed contact binaries to have high obliquities preferentially
between $\sim$40$^{\circ}$ and $\sim$140$^{\circ}$, assuming no
significant obliquity evolution has occurred after formation of the
contact binary. 
Obliquity is defined here as the inclination between
the contact binary's rotational angular momentum vector and the vector
normal to its heliocentric orbital plane. This obliquity angle is
equivalent to the inclination angle defined in Section
\ref{introduction} and used throughout the paper; we assume that the
direction of an initially-detached binary's orbital angular momentum
vector is the same as the direction of a then-collided contact
binary's rotational angular momentum vector.

Observed near-Earth contact binaries include Castalia \citep{huds94},
Bacchus \citep{benn99}, Mithra \citep{broz10}, 1996 HW1
\citep{magr11}, and Itokawa \citep[e.g.][]{ostr04}; all of which have
high obliquities (Busch et al., in prep.). In the inner main belt, small
binaries are observed with obliquities concentrated toward 0$^{\circ}$
and 180$^{\circ}$ and there is a lack of obliquities near 90$^{\circ}$
\citep{prav11}. The concentration of binaries with low inclinations
and the concentration of contact binaries near high inclinations hints
that the Kozai effect may be effective at disrupting high-inclination
binaries, but \citet{prav11} have shown that the Kozai effect cannot
completely explain the observed concentration of binaries with
obliquities near 0$^{\circ}$ and 180$^{\circ}$.

Although the Kozai effect is not responsible for the observed binary
pole concentrations of small main belt binaries near 0$^{\circ}$ and
180$^{\circ}$, it can be effective for more rarely observed NEA
binaries with wide separations.  Our simulations show that
widely-separated binaries can escape the dominating influence of
primary oblateness to be significantly affected by solar
perturbations. As a result, we suggest that wide binaries with
eccentric or highly-inclined orbits may become unstable under the
effects of Kozai, leading to the formation of near-Earth contact
binaries.  Alternate theories of contact binary formation include
low-velocity collisions between components in an unstable binary due
to planetary encounters or radiative effects such as YORP and BYORP
\citep[e.g.][]{tayl11}.

\section{Conclusion} \label{conclusion}

We explored the effect of Kozai cycles caused by the Sun on a sample
of NEA binaries. Kozai oscillations can be suppressed by significant
sources of pericenter precession; we identified 3 processes--primary
oblateness ($J_2$), presence of an additional satellite, and
tides--that contribute to orbital precession.  We determined that primary
oblateness is a dominant source of pericenter precession for small
near-Earth satellites.  Accordingly, we performed numerical
simulations to evaluate the strength of Kozai cycles for NEA binaries
in our sample with the inclusion of the primary's $J_2$.

Our study showed that the binaries in our sample (with the exception
of a hypothetical wide binary modeled after 1998~ST27) do not undergo Kozai cycles due to
their small component separations.  This sample includes all
well-characterized binaries as defined by \citet{fang11ii} but does
not include all observed binaries.  Even the presence of a minimal
$J_2$ (0.001) prevented NEA binaries in our sample from exhibiting
signs of Kozai cycles. Consequently, we conclude that the Kozai effect
is not relevant in explaining the observed characteristics of typical,
observed NEA binaries.  However, we note that the Kozai effect may
have shaped the observed population of binaries by eliminating those
binaries with very low-$J_2$ primaries or widely-separated components.

For rarer observed cases of wide binaries, we studied a hypothetical
wide binary modeled after 1998~ST27 whose large component separation
indicated weaker $J_2$ effects and visible Kozai cycles at a wide
range of possible $J_2$ values. For this hypothetical wide binary, we
were able to map out orbit pole orientations that are stable under
Kozai effects. For cases of such wide binaries, the Kozai effect can
lead to collisions between components.  Accordingly, we suggest that
the Kozai effect acting on widely-separated binaries may be a route to
the formation of near-Earth contact binaries.

\acknowledgments

We thank Simon Porter and the referee, Matija {\'C}uk,
for useful discussions. This work was partially
supported by NASA Planetary Astronomy grant NNX09AQ68G.

\bibliographystyle{apj}
\bibliography{kozai}

\end{document}